\documentclass[lettersize,journal]{IEEEtran}
\usepackage{amsmath,amsfonts}
\usepackage{algorithm}
\usepackage{array}
\usepackage[caption=false,font=normalsize,labelfont=sf,textfont=sf]{subfig}
\usepackage{textcomp}
\usepackage{stfloats}
\usepackage{url}
\usepackage{verbatim}
\usepackage{graphicx}
\usepackage{cite}
\usepackage{algpseudocode}

\usepackage{textcomp,mathcomp}
\usepackage{multirow}
\usepackage{hyperref}
\usepackage{xcolor}
\begin{document}

\title{Fast Thermal-Aware Chiplet Placement Assisted by Surrogate}

\author{Qinqin Zhang, Xiaoyu Liang, Ning Xu and Yu Chen
        % <-this % stops a space
\thanks{Qinqin Zhang, Xiaoyu Liang and Ning Xu are with the School of Information Engineering, Wuhan University of Technology, Wuhan, China.}% (email: qinqinz@whut.edu.cn; liangxiaoyu@whut.edu.cn).}% <-this % stops a space
\thanks{Yu Chen is with the School of Mathematics and Statistics, Wuhan University of Technology, Wuhan, China (email: ychen@whut.edu.cn).}
\thanks{This research is partially supported by the Fundamental Research Funds for the Central University (No. 104972024KFYjc0055), and partially supported by the National Nature Science Foundation of China (No. 92373102).}}

\markboth{IEEE JOURNAL ON EMERGING AND SELECTED TOPICS IN CIRCUITS AND SYSTEMS}%
{Qinqin Zhang, Xiaoyu Liang,
\MakeLowercase{\textit{et al.}}: Fast Thermal-Aware Chiplet Placement Assisted by Surrogate}


\maketitle

\begin{abstract}
With the advent of the post-Moore era, the 2.5-D advanced package is a promising solution to sustain the development of very large-scale integrated circuits. However, the thermal placement of chiplet, due to the high complexity of thermal simulation, is very challenging. In this paper, a surrogate-assisted simulated annealing algorithm is proposed to simultaneously minimize both the wirelength and the maximum temperature of integrated chips. To alleviate the computational cost of thermal simulation, a radial basis function network is introduced to approximate the thermal field, assisted by which the simulated annealing algorithm converges to the better placement in less time.  Numerical results demonstrate that the surrogate-assisted simulated annealing algorithm is competitive to the state-of-the-art thermal placement algorithms of chiplet, suggesting its potential application in the agile design of 2.5D package chip.
% With the advancement of technology and the development of the times, an increasing number of people have started to pay attention to Chiplet technology, and the studies on 2.5D systems have become more and more in-depth. Due to the relatively high requirements of power, it is easy to have thermal problems in 2.5D systems. Therefore, it is very important to do some research on the chiplet placement algorithm to reduce the maximum operating temperature of 2.5D system. In this paper, a chiplet thermal-aware placement algorithm based on simulated annealing algorithm is proposed. In order to obtain better results, the number of iterations in this paper is set to be a large one. However, thermal simulation is a time-consuming process. If we still use the commercial software or the open-source tool Hotspot to solve the temperature, it will consume a great deal of time. So we propose the surrogate-assisted chiplet thermal-aware placement algorithm.
\end{abstract}

\begin{IEEEkeywords}
Chiplet, thermal-aware placement, surrogate, radial basis function network, simulated annealing algorithm
% Chiplet, surrogate-assisted, Radial Basis Function Network, placement algorithm, Simulated Annealing algorithm.
\end{IEEEkeywords}

\section{Introduction}
\IEEEPARstart{T}{he} integrated circuit (IC) has been following Moore's Law in the first decades of its development. However, with the progress of science and technology and the development of the semiconductor industry, the chip process size has approached the physical limit, which results in a decline in chip yield and a rise in chip cost~\cite{ref1}. Therefore, IC has begun to enter the "post-Moore era" guided by non-Moore's law, and Chiplet technology has emerged.

Chiplet technology is a technique that divides a complex monolithic integrated circuit into several smaller chips that can realize sub-functions as required and then integrates these small chips through advanced packaging technology. Nowadays, regarding the packaging forms of chiplets, the most common ones are 2D multi-chip module, 2.5D fan-out wafer-level packaging, 2.5D silicon interposer and 2.5D embedded bridge. The silicon interposer is usually divided into passive interposer and active interposer. There are transistors in the active interposer, and the process technology applied to the active interposer is more complex than that applied to the passive interposer~\cite{ref2}. Therefore, the production cost of the active interposer is higher, and the yield is lower. Considering cost and other issues, the studies focus on chips based on 2.5D passive interposer integration technology.

The implementation of chiplet technology can usually improve the process yield and reduce the cost~\cite{ref3}. Chiplets can be reused in different products to improve the scalability of chips. In general, chiplet has lower combined costs and higher benefits. However, chiplet technology also faces many challenges, such as thermal effects, electromagnetic challenges, capacitive coupling, inductive coupling, mechanical stress and so on. Among them, the influence of thermal effects is the most prominent~\cite{ref1}. With increasing demand for processing large amounts of data, the power consumption of the 2.5D system is increasing, so the possibility of thermal problems in the 2.5D system is also increasing. However, the aim of traditional placement methods is to reduce the total wirelength and minimize the area, so the final results of traditional placement algorithms are usually compact~\cite{ref2}. This also means that it is difficult for traditional placement methods to avoid thermal problems for 2.5D systems with high power density. Therefore, we need to consider how to reduce the maximum operating temperature of 2.5D systems. Due to the high computational complexity of thermal simulation, simultaneous minimization of both wirelength and temperature is time-consuming, making it challenging to design an efficient algorithm for the thermal-ware placement of chiplet.

In this paper, we propose a thermal-aware chiplet placement algorithm assisted by surrogate (TACPAs) to optimize the wirelength and reduce the maximum operating temperature of the 2.5D system. The algorithm is mainly based on simulated annealing algorithm (SA). For the thermal simulation of each placement result in the optimization process, we do not choose commercial thermal simulation software to get the maximum temperature of placement, but make use of the idea of surrogate model to predict the maximum temperature by training a radial basis function network (RBFN). Consequently, the proposed TACPAs can converge to optimized placement results in less time.

This paper is structured as follows.  Related work is reviewed in Section~\ref{sec2}. Section~\ref{sec3} presents the details about the proposed placement algorithm, and Section~\ref{sec4} presents the experimental results with analysis. Funially, Section~\ref{sec5} summarizes this paper.

\section{Related Work}
\label{sec2}
Nowadays, an increasing number of people have started to pay attention to Chiplet. Li \emph{et al.}~\cite{ref4} proposed a reusable general interposer architecture to amortize non-recurring engineering costs and it is capable of significantly speeding up integration flows of interposers across different chiplet-based systems. Zhi Li \emph{et al.}~\cite{ref5} combined the differential evolution (DE) and the particle swarm optimization (PSO) algorithms with the Metropolis rule that can jump out of the local optimum to optimize the impedance of chiplet-based 2.5D integrated circuits. Raikar and Stroobandt~\cite{ref6} developed a modular placement algorithm that has achieved a very good balance between runtime efficiency and placement quality. These studies have carried out optimizations on the placement of the chiplet-based systems from various perspectives, but the thermal effect of chiplet is not considered.

Given the substantial impact of thermal effects on the chiplet-based systems, numerous placement algorithms have emerged recently with the aim of lowering the maximum operating temperature of the chiplet-based systems. Ma \emph{et al.}~\cite{ref2} presented the TAP-2.5D algorithm to jointly minimize the temperature and total wirelength, and it is the first open-source network routing and thermally-aware chiplet placement methodology for heterogeneous 2.5D systems. The SP-CP algorithm proposed in~\cite{ref7} combines the sequence-pair based tree, branch-and-bound method, and advanced placement/pruning techniques. It can find the result fast with the optimized total wirelength. It also proposes a post placement procedure to lower the maximum operating temperature of 2.5D integrated circuits. Duan \emph{et al.}~\cite{ref8} presented an efficient early-stage floorplanning tool named RLPlanner for chiplet-based systems. Making use of the advanced reinforcement learning, the RLPlanner can jointly minimize the wirelength and the maximum operating temperature. Based on the RLPlanner, Deng \emph{et al.}~\cite{ref9} proposed a learning to rank approach with graph representation to choose the best chiplet placement order for each chiplet-based system. Actually, it achieves a $10.05\%$ reduction in wirelength and a $1.01\%$ improvement in peak system temperature. In ~\cite{ref10}, a reinforcement learning (RL) framework is employed to reduce thermal hotspots, improve heat dissipation and enhance system performance.

Some pioneering research has been done on fast thermal analysis of chiplet-based systems. The HotSpot compact thermal modeling approach in ~\cite{ref11} is able to provide detailed static and transient temperature information and it is computationally efficient. Chen \emph{et al.}~\cite{ref12} deployed a novel graph convolutional networks (GCN) architecture to generate the thermal map of the chiplet-based systems and the time it takes to run is less than that of HotSpot. Wang \emph{et al.}~\cite{ref13} introduced a multiscale anisotropic thermal model, which takes the feature-scale thermal conductivities of different materials into account to predict the package-scale steady-state temperature fields. %Unlike these methods of fast thermal analysis, our paper utilizes the surrogate model in the thermal simulation to reduce the running time.

\section{Thermal-Aware Chiplet Placement Assisted by Surrogate}
\label{sec3}
\subsection{Surrogate Approximation of Thermal Map Via the Radial Basis Function Network}
\label{sec3.1}
Commercial thermal simulation software usually outputs the thermal map of the entire 2.5D system by solving the heat conduction equation by finite element method, so it takes a lot of time to use commercial thermal simulation software to simulate the results of placement. The open-source tool HotSpot builds thermal models as lumped thermal RC networks in order to reduce the computation needed~\cite{ref11}. Actually, HotSpot has improved the efficiency of thermal simulation while ensuring the accuracy of simulated temperature. However, when striving for more stable and optimal results, we often increase the number of iterations in the optimization process. As the number of iterations increases, relying solely on HotSpot to obtain the maximum temperature for each placement result during the optimization process still demands a significant amount of time. To achieve better results within a reasonable time frame, this paper proposes the construction of a surrogate model for thermal simulation.

Sometimes the objective functions of optimization problem are not presented in explicit expressions. Instead, it needs to perform numerical simulations or physical experiments to obtain the objective functions, which results in a long time for each evaluation of the objective function~\cite{ref14}. Such optimization problems are commonly referred to as expensive optimization problems. In order to improve efficiency and reduce running time, the concept of surrogate-assisted evolutionary algorithms has been proposed~\cite{ref15}. Surrogate models are typically constructed using historical evaluation data and can serve as approximations of the actual objective functions~\cite{ref16}. By replacing the majority of the time-consuming real function evaluations with surrogate models during the optimization process, significant time savings can be achieved, thereby improving overall efficiency. Common types of surrogate models include polynomial regression model, radial basis function (RBF), Gaussian process, artificial neural network, support vector machine and so on~\cite{ref16}. Due to the simple structure and general approximation ability of RBFN, this paper chooses RBF to build the surrogate model of temperature simulation.

\begin{figure}[!t]
\centering
\includegraphics[width=2.5in]{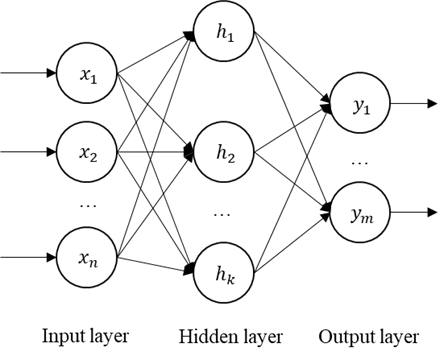}
\caption{The structure diagram of RBFN.}
\label{fig1}
\end{figure}

The structure of RBFN is mainly divided into three layers: the input layer, the hidden layer and the output layer~\cite{ref17}. The structure diagram is shown in Fig.~\ref{fig1}. The input layer is responsible for transmitting the input data to the hidden layer. The hidden layer contains a certain number of RBF neurons, and the number of nodes in the hidden layer is generally determined according to specific problems~\cite{ref18}. The output layer outputs the final result by linearly weighting the weight vector of each node in the hidden layer. In order to build a suitable RBFN, the most important thing is to determine the number of nodes in the hidden layer, the centers and widths of RBF and the connection weights.

In order to simplify the problem, the input layer of our RBFN does not consider the power of the chiplets, but only inputs the widths, heights, center x-coordinates and center y-coordinates of chiplets. Therefore, it is necessary to build its own RBFN for each different case. Assuming the number of training samples is $N$, the number of nodes in the input layer is $n$, and the number of chiplets is $n_{chiplet}$, then the input vector is $X = [X_{1}, X_{2},..., X_{N}]$, and each sample vector is $X_{i} = [x_{i}^1, x_{i}^2,..., x_{i}^n](i\in[1,N], n=4*n_{chiplet})$.

The activation function of RBFN is the radial basis function. Common radial basis functions include Gaussian function, reflected sigmoid function, multiple quadratic function and so on~\cite{ref17}. In this paper, the activation function is taken as the Gaussian function
\begin{equation}
 \label{eq1}
\varphi(X) = \exp\left( - \frac{\left| \left| {X - C_{i}} \right||^{2} \right.}{\sigma^{2}} \right),
 \end{equation}
where $C_{i}(i\in[1,k])$ represents the center of the RBF, $k$ represents the number of nodes in the hidden layer, and $\sigma$ represents the width. In this paper, the centers of the RBF are determined by performing K-means clustering algorithm on the training samples. The data is first clustered into $k$ classes, and then the centers of the classes are taken as the centers of the RBF. Moreover, $\sigma$ is determined according to the distances between the centers of the classes. The number of centers $k$ is set to different values according to the actual situations of different cases. Regarding the solution of the weights between the nodes of the hidden layer and the output layer, the pseudo-inverse method is used in this paper~\cite{ref19}.
In order to reduce the computation burden, the output of the RBFN trained in this paper is only the maximum operating temperature of the 2.5D system, that is, the number of nodes in the output layer $m$ is set to 1. The sample data used in this paper was collected during the operation of the TAP-2.5D algorithm and then processed.

Both the global surrogate model $RBFN_{global}$ and the local surrogate model $RBFN_{local}$ are employed in the proposed algorithm~\cite{ref20}. Samples for training $RBFN_{global}$ should be distributed throughout the decision space to formulate a general bird's-eye view of the landscape, and the local surrogate model $RBFN_{local}$ is trained by data samples surrounding a solution to predict the fitness of the solution more accurately. In this paper, the global surrogate model and the local surrogate model are used alternately to realize the surrogate-assisted chiplet thermal-aware placement algorithm. %The specific application of the surrogate models will be introduced in Section~\ref{sec3.3}.

%{\color{red}Training of the RBFN can be efficiently implemented by addressing a system of linear equations. (here a reference about the RBFN should be cited.)}
To train surrogate models, the selection of training samples is also very important. For $RBFN_{global}$ used in this paper, the samples are sorted in ascending order of temperature, and then a certain number of samples are selected at each temperature as training samples. Note that if the fit at a certain temperature is not good, the training samples at that temperature can be appropriately increased. For $RBFN_{local}$ used in this paper, 500 samples closest to the new solution are selected for training. Then, the temperature simulation can be carried out via Algorithm \ref{alg3}. As the simulated annealing algorithm runs, it becomes more and more likely to use $RBFN_{local}$ to more accurately predict the temperature of the neighbor placement. %{\color{red}brief presentation about the process of ALgorithm \ref{alg3}}
\begin{algorithm}[htp]
  \caption{Temperature Simulation by RBFN.}
  \label{alg3}
  \begin{algorithmic}[1]
    \Require
         the neighbor placement $system_{new}$, the simulated annealing temperature $T_{annealing}$, the global surrogate model $RBFN_{global}$;
     \Ensure
 	the maximum temperature of $system_{new}$ solved by the surrogate model $T_{new}$;
    \If {$T_{annealing}\in (0.2,0.7\rbrack$}
        \State $flag \leftarrow 0.9$;
    \ElsIf{$T_{annealing}\in (0.08,0.2\rbrack$}
        \State $flag \leftarrow 0.8$;
    \ElsIf{$T_{annealing} \leq 0.08$}
        \State $flag \leftarrow 0.7$;
    \Else
        \State $flag \leftarrow 1$;
    \EndIf
    \State Randomly generate $f$ in $\lbrack 0,1)$;
    \If{$flag<f$}
       \State Train the local surrogate model $RBFN_{local}$ based on $system_{new}$;
       \State $T_{new}\leftarrow$ the maximum temperature of $system_{new}$ solved by $RBFN_{local}$;
    \Else
        \State $T_{new}\leftarrow$ the maximum temperature of $system_{new}$ solved by $RBFN_{global}$;
    \EndIf
    \State Return $T_{new}$;
  \end{algorithmic}
\end{algorithm}

\subsection{Wirelength Estimation}
\label{sec3.2}
In this paper, the mixed-integer linear programming (MILP) solver is used to solve the wirelength optimization problem of 2.5D systems. The method was proposed by Coskun \emph{et al.}~\cite{ref21}. In order to simplify the wirelength optimization problem, our paper divides the microbumps required by chiplets into four pin clumps and distributes them around chiplets. The process of routing optimization is to find the routing solution and the microbump assignment to pin clumps that can minimize the interconnects between chiplets while complying with relevant constraints. The successful operation of the MILP solver requires the information of the location of chiplets, the size of chiplets, and the connection between chiplets. The objective function of the routing optimization process is
\begin{equation}
 \label{eq2}
Minimize:{\sum\limits_{i \in C,l \in P,j \in C,k \in P,n \in N_{net}}{d_{iljk} \cdot f_{iljk}^{n}}},
\end{equation}
where $C$ is the set of chiplets, $P$ is the set of pin clumps and $N_{net}$ is the set of nets. $d_{iljk}$ represents the Manhattan Distance from pin clump $l$ on chiplet $i$ to pin clump $k$ on chiplet $j$. $f_{iljk}^n$ represents the number of wires from pin clump $l$ on chiplet $i$ to pin clump $k$ on chiplet $j$ that belong to net $n$.

\subsection{The Thermal-Aware Chiplet Placement Algorithm Assisted by Surrogate}
\label{sec3.3}
%For the placement algorithm, this paper refers to the TAP-2.5D algorithm, the main idea of which is to insert gaps between chiplets to reduce the maximum operating temperature of the entire 2.5D systems.
The proposed placement algorithm follows the general framework of  Tap-2.5D consisting of two stages~\cite{ref2}.
The first stage is to generate the initial placement with optimized wirelength and area using the traditional placement algorithm based on fast simulated annealing algorithm and B* tree~\cite{ref22}. In the second stage, the simulated annealing algorithm is used to strive for the results with reduced maximum operating temperature of the 2.5D systems.

\subsubsection{The cost function to be minimized.}
%In the process of generating the initial placement, we generate the neighbor placement by three operations: rotate, swap and jump.
The cost function for stage one is
\begin{equation}
 \label{eq3}
cost = 0.5 \times \frac{L - L_{min}}{L_{max} - L_{min}} + 0.5 \times \frac{A - A_{min}}{A_{max} - A_{min}},
\end{equation}
where $L$ represents the wirelength and $A$ represents the area. $L_{min}$, $L_{max}$, $A_{min}$ and $A_{max}$ are their respective values of the maximum and the minimum values.  It should be pointed out that the calculation of the wirelength at this stage does not use the MILP solver mentioned in Section~\ref{sec3.2}. Instead, it takes the classical Half-perimeter Wirelength (HPWL) as the estimation of wirelength. The cost function of the second stage is
\begin{equation}
 \label{eq4}
cost = a \times \frac{T - T_{min}}{T_{max} - T_{min}} + (1 - a) \times \frac{L - L_{min}}{L_{max} - L_{min}},
\end{equation}
where $T$ represents the temperature and $L$ represents the wirelength. The weight $a$ is set according to the specific situation. When the maximum operating temperature of the 2.5D systems is lower than $85\tccentigrade$, we generally focus on optimizing the wirelength and the weight $a$ is set to $0$ in this case.
As can be seen from the cost function of the second stage, we need to perform thermal simulations for each placement result to obtain the maximum operating temperature of the current system. In order to improve efficiency, this paper uses the surrogate-assisted chiplet thermal-aware placement algorithm. Algorithm \ref{alg3} describes how to use the global surrogate model and the local surrogate model alternately.

\begin{algorithm}[!htp]
  \caption{Candidate Generation by Moving}
  \label{alg1}
  \begin{algorithmic}[1]
   \Require
        $granularity$, $T_{annealing}$;
     \Ensure
 	$system_{new}$;
    \State $order_{chiplet} \leftarrow$ the random order of the chiplets;
    \For{$i$ in $order_{chiplet}$}
        \State $order_{direction} \leftarrow$ the random order of directions;
        \For{$j$ in $order_{direction}$}
            \If{$T_{annealing}>0.1$}
                \State $p$ is randomly set to 1, 2, 3;
            \Else
                \State $p \leftarrow 1$;
            \EndIf
            \State $d \leftarrow p*granularity$;
            \If{$j=='left'$}
                \If{chiplet $i$ can be moved $d$ $mm$ to the left}
                   \State Return $system_{new}$;
                \EndIf
            \EndIf
            \If{$j=='right'$}
                \If{chiplet $i$ can be moved $d$ $mm$ to the right}
                   \State Return $system_{new}$;
                \EndIf
            \EndIf
            \If{$j=='up'$}
                \If{chiplet $i$ can be moved up by $d$ $mm$}
                   \State Return $system_{new}$;
                \EndIf
            \EndIf
            \If{$j=='down'$}
                \If{chiplet $i$ can be moved down by $d$ $mm$}
                   \State Return $system_{new}$;
                \EndIf
            \EndIf
        \EndFor
    \EndFor
  \end{algorithmic}
\end{algorithm}

\begin{algorithm}[!htp]
  \caption{Candidate Generation by Jumping}
  \label{alg2}
  \begin{algorithmic}[1]
    \Require
        $T_{annealing}$;
    \Ensure
 	$system_{new}$;
    \State $ratio_{jumping} \leftarrow 0.6*T_{annealing}$;
    \If{$ratio_{jumping}<0.1$}
       \State $ratio_{jumping} \leftarrow 0.1$;
    \EndIf
    \State Randomly generate $r$,$r \in \lbrack 0,1)$;
    \If{$ratio_{jumping}>r$}
       \State Generate $system_{new}$ by the jumping operation;
    \Else
        \State Generate $system_{new}$ by the moving operation;
    \EndIf
    \State Return $system_{new}$;
  \end{algorithmic}
\end{algorithm}

\subsubsection{The generation of candidate placements}
In the second stage, we generate the neighbor placement by two operations. First, a chiplet is randomly moved for a certain distance in a certain direction. This moving operation does not account for the rotation of the chiplets, and the moving distance is correlated with the simulated annealing temperature $T_{annealing}$ as shown in Algorithm \ref{alg1}. The second operation is jumping, which means move a chiplet wherever it can be placed and the chiplet is randomly rotated 90 degrees. The probability of jumping is determined by the simulated annealing temperature $T_{annealing}$ as shown in Algorithm \ref{alg2}. In order to explore the solution space in a wider range in the early stage of the SA algorithm, the moving distance and the probability of jumping is set to the larger ones when the simulated annealing temperature $T_{annealing}$ is higher.

\subsubsection{The framework of TACPAs}
In the optimization process, we do not always use the surrogate model to evaluate the maximum operating temperature of the 2.5D system.Instead, we will sometimes use HotSpot to solve the temperature of the system as shown in Algorithm \ref{alg4}. After every five temperature predictions using RBFN, we employ HotSpot to calculate the system temperature. The results obtained from HotSpot are then saved as samples for subsequent updates to the RBFN. In Algorithm \ref{alg4}, we encapsulate the calculation of the temperature and the acceptance probability into a function. The main idea of the surrogate-assisted chiplet thermal-aware placement algorithm is shown in Algorithm \ref{alg5}. It should be pointed out that only when HotSpot is used to calculate the acceptance probability of the neighbor placement and the neighbor placement is accepted, we will determine whether to update the best solution. We will output $system_{best}$ when $T_{annealing}$ is reduced to $T_{min}$ or when HotSpot is used more than 4051 times. The global surrogate model used in our algorithm is updated at regular intervals.

\begin{algorithm}[!htp]
  \caption{Solution of the Acceptance Probability about Candidate}
  \label{alg4}
  \begin{algorithmic}[1]
    \Require
        $system_{new}$, the current placement $system_{current}$, the step of $system_{new}$ $step$, the step of $system_{current}$ $step_{current}$, the wirelength of $system_{current}$ $L_{current}$, the maximum temperature of $system_{current}$ solved by the surrogate model $T_{current}$, $dict_{HotSpot}$, the wirelength of $system_{new}$ solved by MILP $L_{new}$;
     \Ensure
 	the acceptance probability of $system_{new}$ $ap$, the times HotSpot has been performed $count$;
    \State $k\leftarrow 5$;
    \State $count\leftarrow 1$;
    \If {$step$ $\%$ $k==0$}
       \State $T_{newHotspot}\leftarrow$ the maximum temperature of $system_{new}$ solved by HotSpot;
       \State Add $step$ and $T_{newHotspot}$ into $dict_{HotSpot}$;
       \State $count\leftarrow count+1$;
       \State Add $system_{new}$, $T_{newHotspot}$, $L_{new}$ into samples;
       \If{$step_{current}$ in $dict_{HotSpot}$}
          \State $T_{currentHotspot}\leftarrow$ $dict_{HotSpot}[step_{current}]$;
       \Else
            \State $T_{currentHotspot}\leftarrow$ the maximum temperature of $system_{current}$ solved by HotSpot;
            \State Add $step_{current}$ and $T_{currentHotspot}$ into $dict_{HotSpot}$;
            \State $count\leftarrow count+1$;
            \State Add $system_{current}$, $T_{currentHotspot}$, $L_{current}$ into samples;
       \EndIf
       \State $ap\leftarrow$ the acceptance probability of $system_{new}$ calculated by $T_{currentHotspot}$, $L_{current}$, $T_{newHotspot}$ and $L_{new}$;
    \Else
        \State $ap\leftarrow$ the acceptance probability of $system_{new}$ calculated by $T_{current}$, $L_{current}$, $T_{new}$ and $L_{new}$;
    \EndIf
    \State Return $ap$, $count$;
  \end{algorithmic}
\end{algorithm}

\begin{algorithm}[!htp]
  \caption{the Framework of TACPAs}
  \label{alg5}
  \begin{algorithmic}[1]
    \Require
        .cfg file includes the widths, heights, power of chiplets, connection relation matrix between chiplets, the interposer size $size_{intp}$, the SA decay $decay$ and so on;
     \Ensure
 	the best placement $system_{best}$;
    \State Initialize an empty dictionary $dict_{HotSpot}$;
    \State $counter\leftarrow 0$;
    \State $system_{current}\leftarrow $generate an initial placement using the algorithm based on fast-SA and B* tree;
    \State Train $model_{global}$;
    \State $step,step_{current}\leftarrow 1$;
    \State $T_{current}\leftarrow$ the maximum temperature of the initial placement solved by $model_{global}$;
    \State $T_{best}\leftarrow$ the maximum temperature of the initial placement solved by HotSpot;
    \State Add $step$ and $T_{best}$ into $dict_{HotSpot}$;
    \State $L_{current}, L_{best}\leftarrow$ the wirelength of the initial placement solved by MILP;
    \State $T_{annealing}\leftarrow 1, T_{min}=0.00001$;
    \While{$T_{annealing}>T_{min}$}
          \State $i\leftarrow 1$;
          \While{$i\leq size_{intp}$}
                 \State $step\leftarrow step+1$
                 \State $system_{new}\leftarrow$ generate the neighbor placement;
                 \State $T_{new}\leftarrow$ solve the temperature by the surrogate model;
                 \State $L_{new}\leftarrow$ the wirelength of $system_{new}$ solved by MILP;
                 \State $ap\leftarrow$ solve the acceptance probability of the neighbor placement;
                 \State Randomly generate $a$,$a \in \lbrack 0,1)$;
                 \If{$ap>a$}
                    \State $system_{current}\leftarrow$ $system_{new}$, $T_{current}\leftarrow$ $T_{new}$, $L_{current}\leftarrow$ $L_{new}$, $step_{current}\leftarrow$ $step$;
                    \If {$step$ $\%$ $k==0$}
                        \State $bap\leftarrow$ the acceptance probability calculated by $T_{best}$, $L_{best}$, $T_{newHotSpot}$ and $L_{current}$;
                        \If{$bap>1$}
                           \State $system_{best}\leftarrow$ $system_{current}$,  $T_{best}\leftarrow$ $T_{newHotSpot}$, $L_{best}\leftarrow$ $L_{current}$;
                        \EndIf
                    \EndIf
                \EndIf
                \If{$count\geq 4051$}
                   \State Return $system_{best}$
                \EndIf
                \State $i\leftarrow i+1$;
          \EndWhile
          \State $T_{annealing}\leftarrow T_{annealing}*decay$;
          \State $counter\leftarrow counter+1$
          \If{$counter\geq 10$}
              \State $model_{global}\leftarrow$ update the global surrogate model
          \EndIf
    \EndWhile
    \State Return $system_{best}$;
  \end{algorithmic}
\end{algorithm}

\section{Numerical Results}
\label{sec4}
In this section, the results of the proposed TACPAs are compared with the traditional placement algorithm which generates the initial placement of our algorithm, the TAP-2.5D algorithm and the RLPlanner algorithm. All the cases used in this paper come from the open source of~\cite{ref2}. In the proposed TACPAs, the simulated annealing temperature  $T_{annealing}$ decays with a factor of 0.97. The $granularity$ in this paper is $1mm$ and the size of the interposer $size_{intp}$ is set to $45mm$. It should be noted that the steps of our method are more than twice that of TAP-2.5D algorithm, but the running time of our algorithm is only $10-20$ minutes slower than TAP-2.5D algorithm. %Our computer has a NVIDIA GeForce RTX 4090.

\subsection{Case 1: Multi-GPU system}
\label{sec4.1}

\begin{figure*}[!htp]
\centering
  \subfloat[Compact placement algorithm]{\includegraphics[width=0.33\textwidth]{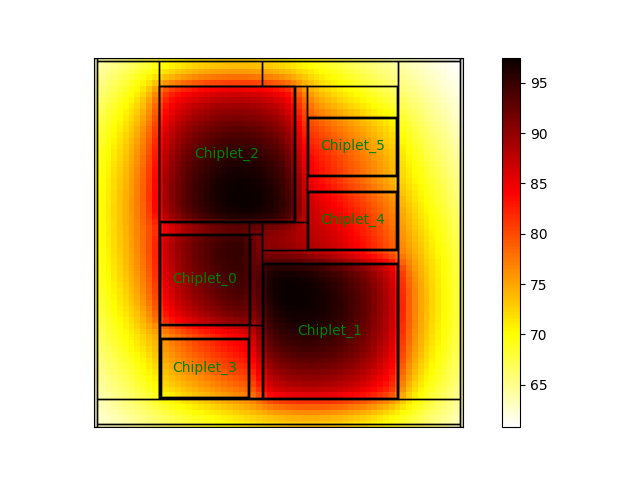}
      \label{fig2}}
\subfloat[TAP-2.5D]{\includegraphics[width=0.33\textwidth]{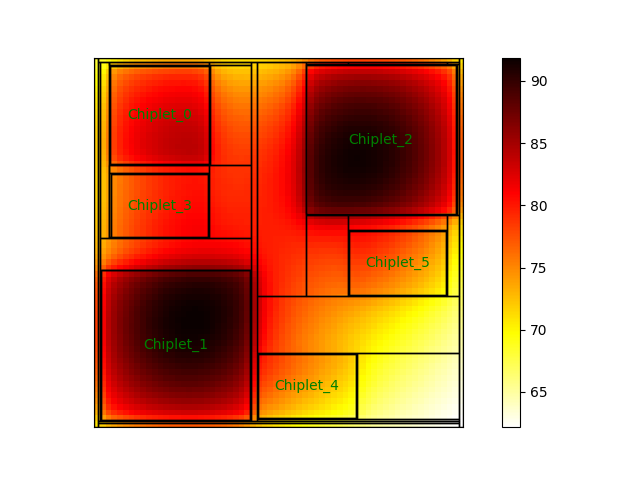}\label{fig3}}
\subfloat[TACPAs]{\includegraphics[width=0.33\textwidth]{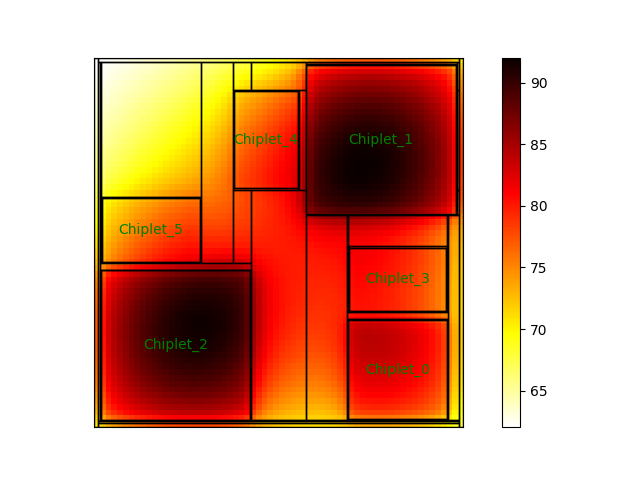}\label{fig4}}
\caption{The thermal maps of the final results for the multi-GPU system.}\label{fig:multiGPU}
\end{figure*}

% \begin{figure}[!t]
% \centering
% \includegraphics[width=3in]{multigpu_compact.png}
% \caption{The thermal map of the final result obtained by using the classical compact placement algorithm for the Multi-GPU System.}
% \label{fig2}
% \end{figure}

% \begin{figure}[!t]
% \centering
% \includegraphics[width=3in]{multigpu_TAP.png}
% \caption{The thermal map of the final result obtained by using the TAP-2.5D algorithm for the Multi-GPU System.}
% \label{fig3}
% \end{figure}

% \begin{figure}[!t]
% \centering
% \includegraphics[width=3in]{multigpu_TACPAs.png}
% \caption{The thermal map of the final result obtained by using the TACPAs algorithm for the Multi-GPU System.}
% \label{fig4}
% \end{figure}

The thermal maps of the multi-GPU system are illustrated in Fig. \ref{fig:multiGPU}. Fig.~\ref{fig2} shows the final result of the Multi-GPU System, which is obtained by using the traditional algorithm. The temperature of the result is $95.27\tccentigrade$ and the wirelength is $88093mm$. The result of the TAP-2.5D algorithm is shown in Fig.~\ref{fig3}. It has a lower temperature of $91.24\tccentigrade$ but a longer wirelength of $93389mm$. Fig.~\ref{fig4} shows the result obtained by using the proposed algorithm. Our TACPAs produces a temperature of $91.33\tccentigrade$ but the wirelength is $86079mm$. Compared with the traditional algorithm, our algorithm can reduce the system temperature. Table~\ref{tab1} shows the results of the TAP-2.5D algorithm, the RLPlanner algorithm and the proposed TACPAs algorithm for the Multi-GPU System. Our algorithm results in a temperature $0.09\tccentigrade$ higher than that of TAP-2.5D but $7310mm$ less wirelength than that of TAP-2.5D. Compared with the RLPlanner algorithm, the temperature of our TACPAs is $0.18\tccentigrade$ higher but the wirelength is $11663mm$ less.

\begin{table}
\begin{center}
\caption{COMPARISON BETWEEN DIFFERENT ALGORITHMS FOR MULTI-GPU SYSTEM}
\label{tab1}
\begin{tabular}{| c | c | c |}
\hline
\multirow{2}{*}{Method} & \multicolumn{2}{c|}{Multi-GPU System} \\
\cline{2-3}
 & Temperature $/ \tccentigrade$ &
Wirelength $/mm$\\
\hline
TAP-2.5D & $91.24$ & $93389$\\
\hline
RLPlanner & $91.15$ & $97742$\\
\hline
TACPAs & $91.33$ & $86079$\\
\hline
\end{tabular}
\end{center}
\end{table}

\subsection{Case 2: CPU-DRAM system}
\label{sec4.2}

\begin{figure*}[!htp]
\centering
  \subfloat[Compact placement algorithm]{\includegraphics[width=0.33\textwidth]{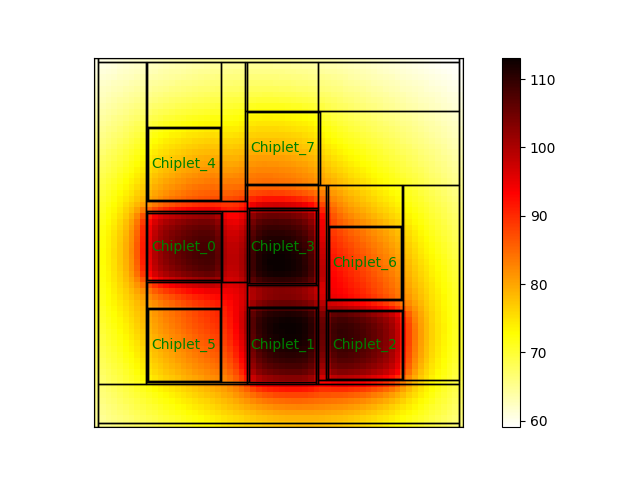}
      \label{fig5}}
\subfloat[TAP-2.5D]{\includegraphics[width=0.33\textwidth]{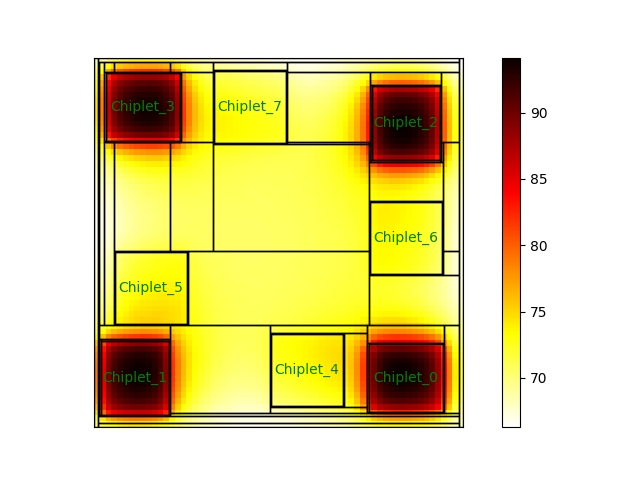}\label{fig6}}
\subfloat[TACPAs]{\includegraphics[width=0.33\textwidth]{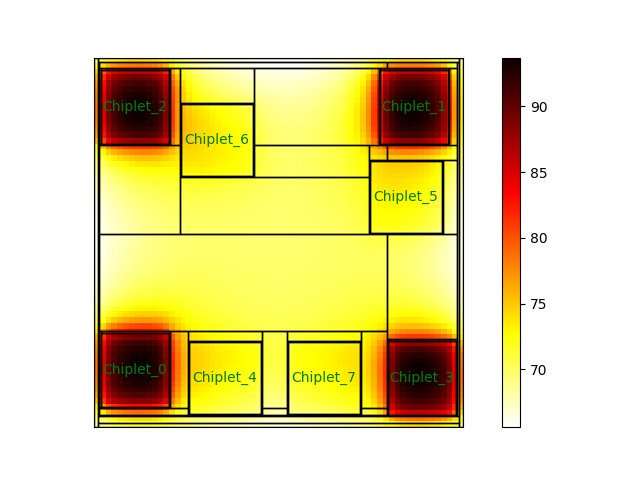}\label{fig7}}
\caption{The thermal maps of the final results for the CPU-DRAM system~\cite{ref23}.}\label{fig:CPUDRAM}
\end{figure*}

% \begin{figure}[!t]
% \centering
% \includegraphics[width=3in]{micro150_compact.png}
% \caption{The thermal map of the final result obtained by using the classical compact placement algorithm for the CPU-DRAM System.}
% \label{fig5}
% \end{figure}

% \begin{figure}[!t]
% \centering
% \includegraphics[width=3in]{micro150_TAP.png}
% \caption{The thermal map of the final result obtained by using the TAP-2.5D algorithm for the CPU-DRAM System.}
% \label{fig6}
% \end{figure}

% \begin{figure}[!t]
% \centering
% \includegraphics[width=3in]{micro150_TACPAs.png}
% \caption{The thermal map of the final result obtained by using the TACPAs algorithm for the CPU-DRAM System.}
% \label{fig7}
% \end{figure}

\begin{table}
\begin{center}
\caption{COMPARISON BETWEEN DIFFERENT ALGORITHMS FOR CPU-DRAM System}
\label{tab2}
\begin{tabular}{| c | c | c |}
\hline
\multirow{2}{*}{Method} & \multicolumn{2}{c|}{CPU-DRAM System} \\
\cline{2-3}
 & Temperature $/ \tccentigrade$ &
Wirelength $/mm$\\
\hline
TAP-2.5D & $94.15$ & $169716$\\
\hline
RLPlanner & $92.88$ & $176246$\\
\hline
TACPAs & $93.63$ & $193696$\\
\hline
\end{tabular}
\end{center}
\end{table}

The thermal maps of the multi-GPU system are illustrated in Fig. \ref{fig:CPUDRAM}.
The thermal map of the result based on the traditional algorithm is shown in Fig.~\ref{fig5}. The temperature of the result is $112.39\tccentigrade$ and the wirelength is $103085mm$. The solution of the TAP-2.5D algorithm is shown in Fig.~\ref{fig6}. It produces a lower temperature of $94.15\tccentigrade$ but a longer wirelength of $169716mm$. The result of our TACPAs in Fig.~\ref{fig7} produces a temperature of $93.63\tccentigrade$ and a wirelength of $193696mm$. Obviously, the result of our algorithm has reduced the maximum operating temperature of the 2.5D system compared to the traditional algorithm. Table~\ref{tab2} shows the results of the TAP-2.5D algorithm, the RLPlanner algorithm and the proposed TACPAs algorithm for the CPU-DRAM System. Compared with the TAP-2.5D algorithm, the wirelength obtained by our TACPAs is $23980mm$ longer but the temperature is $0.52\tccentigrade$ lower. Because our algorithm focuses on reducing the temperature when the temperature of the 2.5D system is high.

\subsection{Case 3: Huawei Ascend 910 system}
\label{sec4.3}

\begin{figure*}[!htp]
\centering
  \subfloat[Compact placement algorithm]{\includegraphics[width=0.33\textwidth]{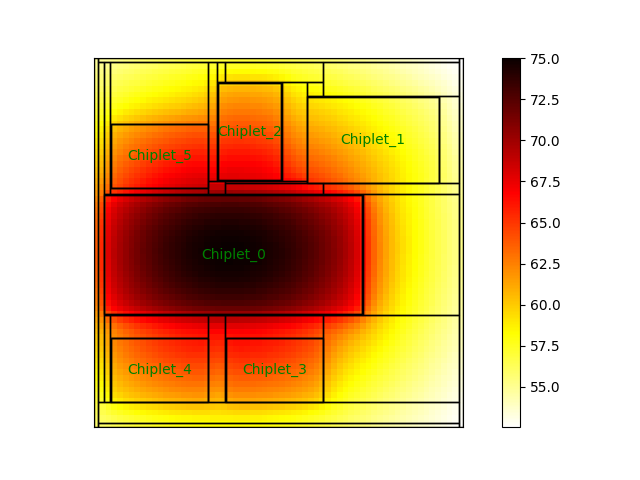}
      \label{fig8}}
\subfloat[TAP-2.5D]{\includegraphics[width=0.33\textwidth]{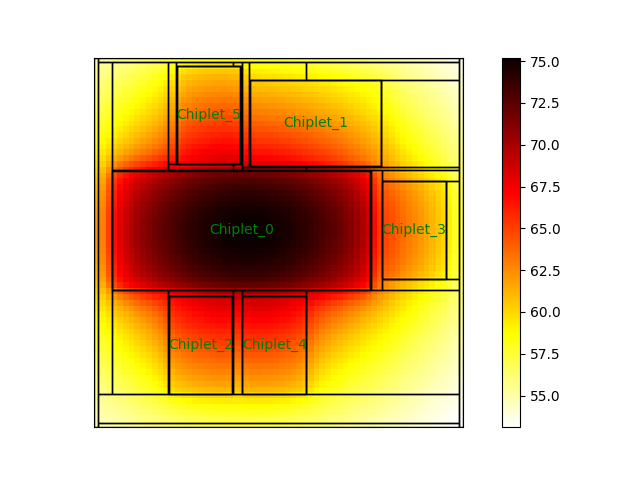}\label{fig9}}
\subfloat[TACPAs]{\includegraphics[width=0.33\textwidth]{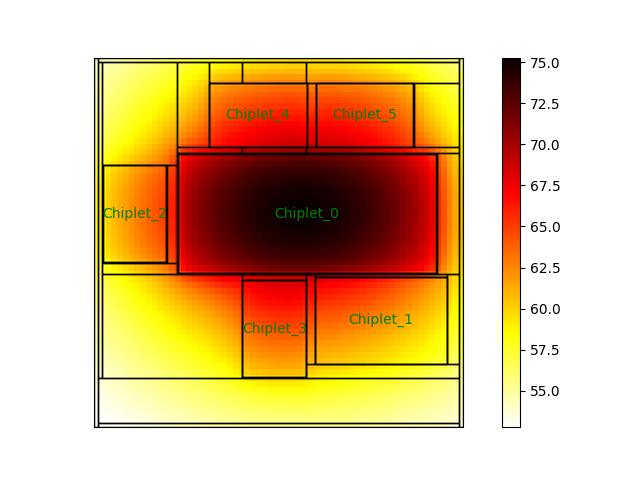}\label{fig10}}
\caption{The thermal maps of the final results for the Huawei Ascend 910 system~\cite{ref24}.}\label{fig:Ascend}
\end{figure*}

% \begin{figure}[!t]
% \centering
% \includegraphics[width=3in]{ascend910_compact.png}
% \caption{The thermal map of the final result obtained by using the classical compact placement algorithm for the Huawei Ascend 910 System.}
% \label{fig8}
% \end{figure}

% \begin{figure}[!t]
% \centering
% \includegraphics[width=3in]{ascend910_TAP.png}
% \caption{The thermal map of the final result obtained by using the TAP-2.5D algorithm for the Huawei Ascend 910 System.}
% \label{fig9}
% \end{figure}

% \begin{figure}[!t]
% \centering
% \includegraphics[width=3in]{ascend910_TACPAs.png}
% \caption{The thermal map of the final result obtained by using the TACPAs algorithm for the Huawei Ascend 910 System.}
% \label{fig10}
% \end{figure}

\begin{table}
\begin{center}
\caption{COMPARISON BETWEEN DIFFERENT ALGORITHMS FOR Huawei Ascend 910 System}
\label{tab3}
\begin{tabular}{| c | c | c |}
\hline
\multirow{2}{*}{Method} & \multicolumn{2}{c|}{Huawei Ascend 910 System} \\
\cline{2-3}
 & Temperature $/ \tccentigrade$ &
Wirelength $/mm$\\
\hline
TAP-2.5D & $75.2$ & $17497$\\
\hline
RLPlanner & $77.12$ & $18130$\\
\hline
TACPAs & $75.26$ & $17350$\\
\hline
\end{tabular}
\end{center}
\end{table}
The thermal maps of the Huawei Ascend 910 System are illustrated in Fig. \ref{fig:Ascend}.
Fig.~\ref{fig8} is the output from the traditional placement algorithm. The temperature is $75.03\tccentigrade$ and the wirelength is $23805mm$. The solution of the TAP-2.5D algorithm shown in Fig.~\ref{fig9} has a temperature of $75.2\tccentigrade$ and the wirelength is $17497mm$. The thermal map of the result obtained by the TACPAs algorithm is shown in Fig.~\ref{fig10}. It produces a temperature of $75.26\tccentigrade$ with a wirelength of $17350mm$. When the temperature is below $85\tccentigrade$, the TAP-2.5D algorithm and the TACPAs algorithm will focus on optimizing the wirelength. Table~\ref{tab3} shows the results of the TAP-2.5D algorithm, the RLPlanner algorithm and the proposed TACPAs algorithm for the Huawei Ascend 910 System. Our TACPAs produces a temperature $0.06\tccentigrade$ higher than that of TAP-2.5D but the wirelength is $147mm$ less than that of TAP-2.5D. Compared with the RLPlanner algorithm, the temperature of our TACPAs is $1.86\tccentigrade$ lower and the wirelength is $780mm$ less. In this case, our algorithm performs better than the others.

\section{Conclusion}
\label{sec5}
In this paper, we propose a surrogate-assisted chiplet thermal-aware placement algorithm (TACPAs) to optimize the wirelength and reduce the maximum operating temperature of the 2.5D systems. The radial basis function network is trained as the global and the local surrogate models, by which lower complexity of thermal simulation is available. Numerical results demonstrate that with a minor increase in the maximum temperature, the proposed TACPAs can further optimize the total wirelength of chiplet systems. The approximate thermal simulation implemented by the radial basis function network significantly reduces the time complexity, suggesting its potential application in the design of efficient thermmal placement algorithm.%In order to obtain better results, we always increase the steps of the optimization process. However, the thermal simulation is a time-consuming process. With the increase of the steps, the running time will also increase.
%Due to the deployment of the global surrogate model and the local surrogate model, the proposed algorithm can achieve competitive placement results without adding a lot of running time.

% \section*{Acknowledgments}
% This research is partially supported by the Fundamental Research Funds for the Central University (No. 104972024KFYjc0055), and partially supported by the National Nature Science Foundation of China (No. 92373102).

\vfill

\end{document}